\documentclass{aa}  
\usepackage[normalem]{ulem}

\usepackage{array}
\usepackage{graphicx}
\usepackage{txfonts}
\usepackage[colorlinks=true,linkcolor=blue, citecolor=blue]{hyperref}
\begin{document}

    \title{Enceladus and Jupiter as exoplanets: the opposition surge effect}

   \author{K. D. Jones\inst{1} \and
          B. M. Morris \inst{2} \and
          K. Heng \inst{3, 4, 5}
          }

\institute{
Center for Space and Habitability, University of Bern, Gesellschaftsstrasse 6, CH-3012 Bern, Switzerland \and
Space Telescope Science Institute, 3700 San Martin Dr, Baltimore, MD 21218, USA \and
Ludwig Maximilian University, University Observatory Munich, Scheinerstrasse 1, Munich D-81679, Germany \and
University of Warwick, Department of Physics, Astronomy \& Astrophysics Group, Coventry CV4 7AL, United Kingdom \and
University of Bern, ARTORG Center for Biomedical Engineering Research, Murtenstrasse 50, CH-3008, Bern, Switzerland 
}

   \authorrunning{K. D. Jones et al.}
 
   \date{Submitted 2024}
   \abstract
   {Planets and moons in our Solar System have strongly peaked reflected light phase curves at opposition. In this work, we produce a modified reflected light phase curve model and use it to fit the Cassini phase curves of Jupiter and Enceladus. This `opposition effect' is caused by shadow hiding (SH; particles or rough terrain cast shadows which are not seen at zero phase) and coherent backscattering (CB; incoming light constructively interferes with outgoing light). We find tentative evidence for CB preference in Jupiter compared to SH, and no evidence of preference in Enceladus. We show that the full-width half-maximum (FWHM) of Jupiter's opposition peak is an order of magnitude larger than that of Enceladus and conclude that this could be used as a solid-surface indicator for exoplanets. We investigate this and show that modelling the opposition peak FWHM in solid-surface exoplanets would be unfeasible with JWST or the Future Habitable Worlds Observatory due to the very large signal-to-noise required over a small phase range. }


   \maketitle

\section{Introduction}
For over a century, observations have shown that phase curves of many Solar System bodies have sharp peaks in reflectance at opposition. It has also been shown that the moons and rocky planets produce narrower peaks than the gas giants \citep[see, e.g.][]{Sudarsky2005, Dyudina2016}. 

Two underlying mechanisms driving the `opposition effect' are: shadow hiding (SH) and coherent backscattering (CB). Shadow hiding, first theorised in \cite{vonSeeliger1887}, occurs on rough surfaces, for example, a rocky planet or moon. When illuminated near quadrature, the roughness casts shadows, reducing the illuminated area and the total reflected light. Close to zero phase, the particles or rough terrain hide their own shadow and the body gets much brighter when viewed face-on. As this involves the shadows cast by the incoming light, this effect only works with singly-scattered light. The second mechanism is coherent backscattering. This is an effect that works with both singly and multiply scattered light. As described thoroughly in \cite{Hapke2002}, CB occurs when the light is scattered in such a way that it coherently interferes with the incoming light and therefore the observer views an increase in brightness near zero phase. This effect can only occur in an inhomogeneous particulate medium where the particles have very high single-scattering albedos (ratio of scattering to absorption) and the particles have radii similar to the wavelength of the incoming light \citep{Helfenstein1997}.

For exoplanets, observing a phase curve is a crucial tool in characterising a planetary atmosphere. With new space and ground-based telescopes coming online, we can begin to probe smaller, more Earth-like planets. These cooler planets require us to observe in the optical to obtain a reflected-light phase curve. Due to all the possible signals which contribute to the overall phase curve, modelling these observations can be difficult and significant inhomogeneities can be detected \cite[see, e.g.][]{Morris2024}. Therefore understanding the signals within phase curves and being able to model them correctly is crucial for the investigations of planetary atmospheres and origins.   

In this paper we investigate whether the opposition effect could be detectable in exoplanet phase curves and what this would reveal about the surface and/or the atmosphere of the planet. In Section \ref{sec:theory}, we develop our own opposition effect phase curve model and test it with multi-wavelength phase curves of Jupiter and Enceladus (see Section \ref{sec:solarsystem}). In Section \ref{sec:fwhm}, we investigate the differences between the full-width half-maximum of the opposition peak in Jupiter and Enceladus. In Section \ref{sec:rockyexoplanets} we look at whether this feature could be used as a solid-surface exoplanet detector. We detail the caveats and further interpretations in Section \ref{sec:discussion}.

\section{Methods}
\subsection{Jupiter and Enceladus Cassini data}
\label{sec:data}
\begin{figure}
   \centering
   \includegraphics[width=9cm]{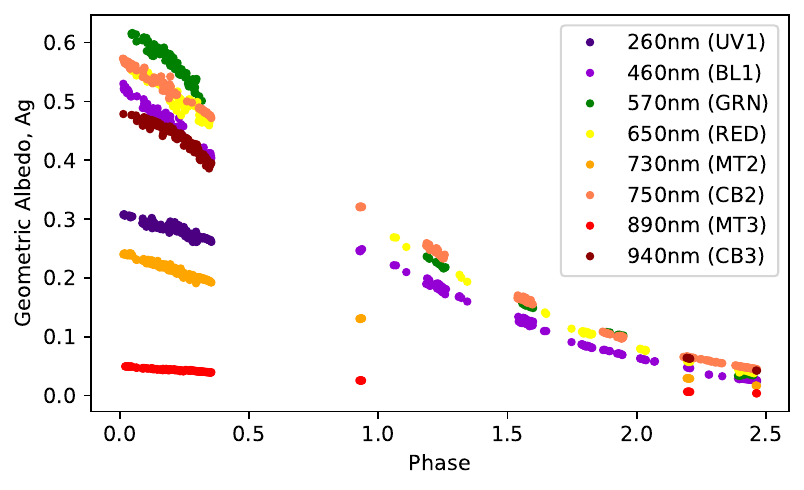}
      \caption{Jupiter Cassini phase curves from \cite{Li2018}, with their corresponding colour filter used in this analysis. The datapoint colours are a crude guide for the different wavelengths of the filters.}
         \label{fig:jup_data}
\end{figure}

\begin{figure}
   \centering
   \includegraphics[width=9cm]{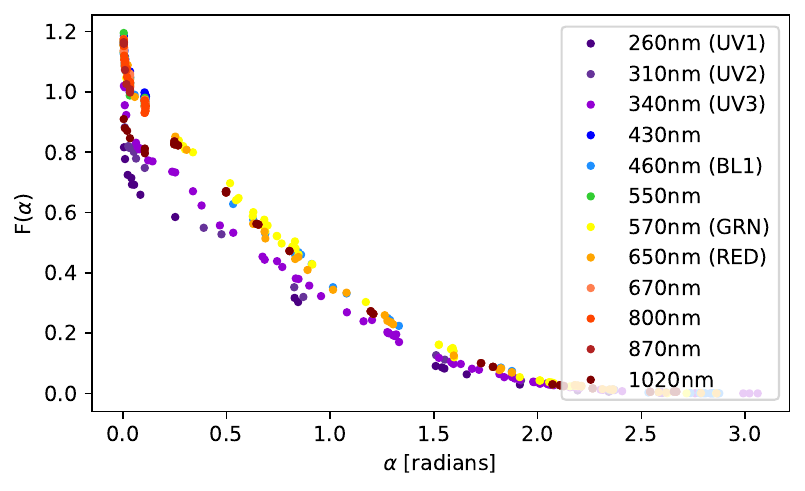}
      \caption{Enceladus Cassini phase curves from \cite{Li2023}, with their corresponding colour filter used in this analysis. The datapoint colours are a crude guide for the different wavelengths of the filters. Some of the filters do not have alternative names.}
         \label{fig:enc_data}
\end{figure}

We use multi-wavelength phase curves of both Jupiter and Enceladus previously published by \cite{Li2018} and \cite{Li2023}, which reported images and combined reflectance phase curves taken with Cassini/ISS, Cassini/VIMS, and the Hubble Space Telescope (HST). Figure \ref{fig:jup_data} shows these Jupiter phase curves as a function of wavelength. Due to the limited number of images taken by the Cassini flyby, the phase coverage is not homogeneous. We limit the dataset to Jupiter phase curves with sufficient phase coverage by removing phase curves with fewer than 40 datapoints. Figure \ref{fig:enc_data} shows the Enceladus phase curves. By eye it is possible to detect, particularly for Enceladus, the sharp increase in the reflected light flux near zero phase. 

\subsection{Reflected light phase curve model}
\label{sec:theory}
To fit these phase curves, we use the reflected light phase curve model developed in \cite{Heng2021b}, and recently applied to more planets in \citet{Morris2024}. This is a unique model that can produce closed-form phase curve solutions for any scattering phase function specified by the user, making it flexible enough for both exoplanets and Solar System objects. The model is analytic and has few free parameters, allowing for fast computing time during model-fitting within a Bayesian inference framework.

Following \cite{Heng2021a}, we model reflected light phase curves with a single Henyey-Greenstein scattering phase function:
\begin{equation}
P(\alpha) = \frac{1-g^2}{(1 + g^2 + 2g\cos{\alpha})^{3/2}}
\label{eqn:shg}
\end{equation}
where $g$ is the scattering asymmetry parameter, and $\alpha$ is the orbital phase angle. This phase function allows for both forward and backward scattering in variable amounts and limits the number of free parameters in the model. 

The reflected light phase curve model describes the reflected flux as a function of phase $F(\alpha)$. We introduce an opposition peak following \citet{Hapke2002} by multiplying an additional term $(1 + \delta_{\textrm{opp}})$ so the reflected flux is: 
\begin{equation}
F = F_\star A_g\Psi(\alpha) \left(1 + \delta_{\textrm{opp}}(\alpha)\right),
\end{equation}
where $F_\star$ is the stellar flux, $A_g$ is the geometric albedo, $\Psi$ is the integral phase function and $\delta_{\rm opp}(\alpha)$ depends on the mechanism for the opposition surge.

There are two possible origins for the opposition peak: shadow hiding (SH) and coherent backscattering (CB). \cite{Hapke2002} notes that the correction term associated with SH should act only on singly scattered light, whereas CB acts on both singly and multiply scattering light (caveat: high levels of multiply scattered light can dilute shadows, which can break down this assumption.) However, \cite{Hapke2002} also notes the similarity of the SH and CB formulae and recommends that the entire opposition effect be modelled by the SH formula alone. For this reason, we apply the correction term to both singly and multiply scattered light for both SH and CB. We aim to investigate the differences between these two models at an appropriate level of accuracy as set by the precision of the data themselves.

From the empirical descriptions of \cite{Hapke1986} and \cite{Hapke1993} we take the SH functional form to be
\begin{equation}
\delta_{\textrm{SH}}(\alpha) = \frac{B_0}{1 + (1/h_{\textrm{SH}})\tan{(\alpha/2)}}
\end{equation}
where $\alpha$ is the phase angle, $B_0$ is a constant describing the amplitude of the SH peak and $h_{\textrm{SH}}$ is a dimensionless constant describing the width of the SH peak. It follows that the full-width half-maximum (FWHM) of the shadow hiding peak is
\begin{equation}
\textrm{FWHM}_{\textrm{SH}} = 2h_{\textrm{SH}}.
\label{eqn:fwhm}
\end{equation}

The form for CB is similar to and derives from \cite{Akkermans1988} with a random walk of photons in a scattering medium, and \cite{Hapke2002}:
\begin{equation}
\delta_{\textrm{CB}}(\alpha) = \frac{B_0}{2}\frac{1 + \frac{1-e^{-x}}{x}}{(1+x)^2}, 
\end{equation}
where
\begin{equation}
x = \frac{\tan(\alpha/2)}{h_{\textrm{CB}}}.
\end{equation}

We expect that SH and CB act on the phase curve together, but it is still uncertain how both effects act simultaneously. For decades, the underlying mechanism of the opposition effect on the Moon was thought to be solely shadow-hiding \citep[see, e.g.][]{Hapke1963, Gehrels1964}. However, by measuring the polarisation of light reflected off lunar soil samples, \cite{Hapke1993} showed that CB was also a major contributor to this effect \citep[with later support in][]{Helfenstein1997, Hapke1998}. Therefore, whilst there is evidence for both mechanisms causing the peak in rocky bodies, it is still not understood which dominates or whether they are also both present to cause the opposition peaks seen in Jupiter and Saturn.

In this paper, we perform a comprehensive model comparison between the SH model, CB model and a non-opposition phase curve model for both Jupiter and Enceladus. We compare models with leave-one-out (LOO) cross validation statistic \citep{Vehtari2015} and interpret our results in Section \ref{sec:modelcomp}. To perform the model fitting for the phase curves, we use \texttt{numpyro} \citep{phan2019composable, bingham2019pyro} with an MCMC sampler running with 8 chains, 2000 burn-in steps and 3000 steps. We confirmed the chains had converged after the fitting procedure by inspecting the Gelman-Rubin statistic $\hat{r} < 1.01$ for each free parameter \citep{Gelman1992}. Tables \ref{tab:bigtable_jup} and \ref{tab:bigtable_enc} shows the priors used for the free parameters within the SH model, which is the same as the priors we used for the CB model. We then measure and compare the FWHM of the opposition peaks in the Jupiter and Enceladus phase curves (see Section \ref{sec:fwhm}).

\section{Solar System results}
\label{sec:solarsystem}
\subsection{Cross-validation favours models with opposition peaks}
\label{sec:modelcomp}
\begin{figure}
   \centering
   \includegraphics[width=9cm]{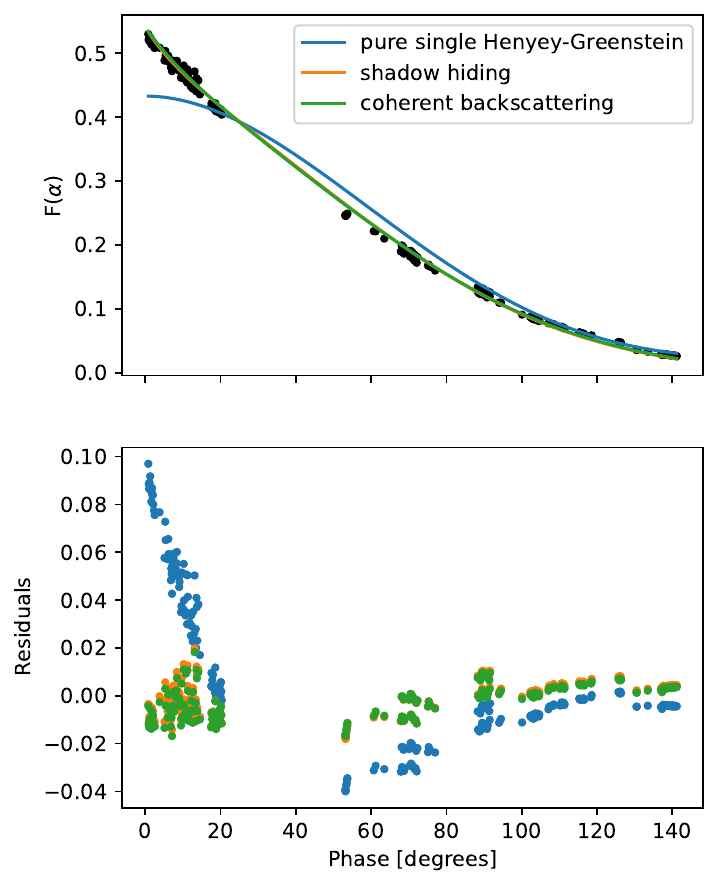}
      \caption{Best fit models to one of Jupiter's phase curves (filter BL1, 463\,nm) for three different models: a pure single Henyey-Greenstein model with no additional opposition peak, shadow hiding model, and coherent backscattering model. Top panel shows the three fits and the second panel shows the residuals of these best fits. It is clear that the model with no opposition peak is the worst fit to the data. The other two models produce, by eye, almost identical fits, which is expected since the functional forms are similar. However the residuals still show subtle differences. Using the LOO model selection statistic, we conclude that the coherent backscattering model is the best fitting model for this phase curve, along with all-but-one of the other Jupiter Cassini phase curves. The corner plot showing the posteriors for the shadow hiding model in this plot are in Appendix \ref{sec:bestfitresultsappendix}, Figure \ref{fig:jup_cornerplot}.}
        \label{fig:bestfitmodels}
\end{figure}

We fit the Jupiter phase curves with an SH model, a CB model and without any opposition model, and compare models with cross validation. The non-opposition effect model was immediately ruled out with no LOO model weight compared to the models with the opposition effect included. Therefore Jupiter likely has an opposition peak, in agreement with previous works including \cite{Heng2021a}, \cite{Dyudina2016} and \cite{Mayorga2016}.  Figure \ref{fig:bestfitmodels} shows an example of the models fitted to the 463\,nm phase curve. Even by visual inspection, it is clear that the model without an opposition peak does not accurately capture the phase curve flux less than around 20 degrees.

Fitting separate CB and SH models to Jupiter's phase curves, we found that 7 out of 8 phase curves prefer the CB model. If this is implying that CB dominates the opposition peak of Jupiter, this could be due to the lack of a surface on Jupiter, preventing dark shadows from forming (and consequently disappearing at opposition). The presence of CB implies multiple-scattering is occurring in Jupiter's atmosphere and that the scattering particles have an inhomogeneous size distribution \citep{Hapke2002, Hapke1998}. 

\begin{figure}
   \centering
   \includegraphics[width=9cm]{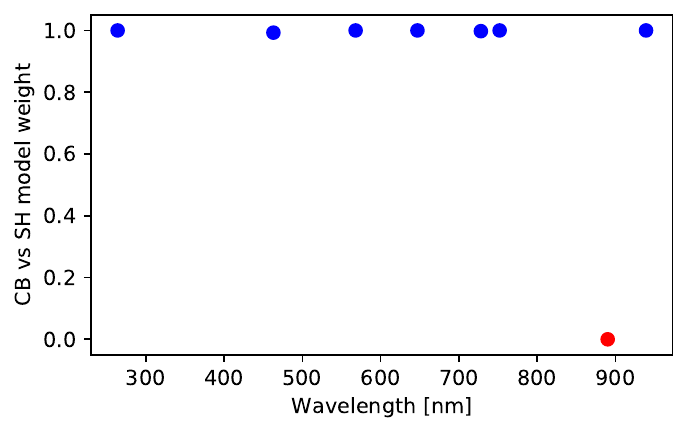}
      \caption{LOO model weight of the CB model vs the SH model for the Jupiter phase curves plotted against wavelength. When the model weight is close to 1, then the CB model is preferred (blue points), however close to 0 indicates that SH is preferred (red points). The red point shows the only phase curve where the SH model is preferred.}
         \label{fig:jup_wavelengthvselpddiff}
\end{figure}

We show the values of the model weights as a function of wavelength in Figure \ref{fig:jup_wavelengthvselpddiff} and in Table \ref{tab:jup_modelselect}. We do not find a trend in wavelength.

\begin{table}
\centering
\caption{Jupiter phase curves' preferred models}
\begin{tabular}{ |c|c|c|c| } 
 \hline
 Wavelength [nm] & Filter & Preferred Model & Weight\\
 \hline
 260 & UV1 &  CB & 1.00 \\ 
 460 & BL1 & CB & 0.99\\ 
 570 & GRN & CB & 1.00\\ 
 650 & RED & CB& 1.00\\
 730 & MT2 &  CB& 1.00\\
 750 & CB2 & CB& 1.00\\
 890 & MT3 & SH& 1.00\\
 940 & CB3 & CB& 1.00\\
 \hline
\end{tabular}
\tablefoot{For each of Jupiter's Cassini phase curves we report here the preferred model from our fitting procedure. 7/8 phase curves prefer the coherent backscattering model (CB) over the shadow hiding model (SH). The wavelength column refers to the effective wavelength of the filter used. `Weight' is the LOO model weight of the preferred model.}
\label{tab:jup_modelselect}
\end{table}

\begin{figure}
   \centering
   \includegraphics[width=9cm]{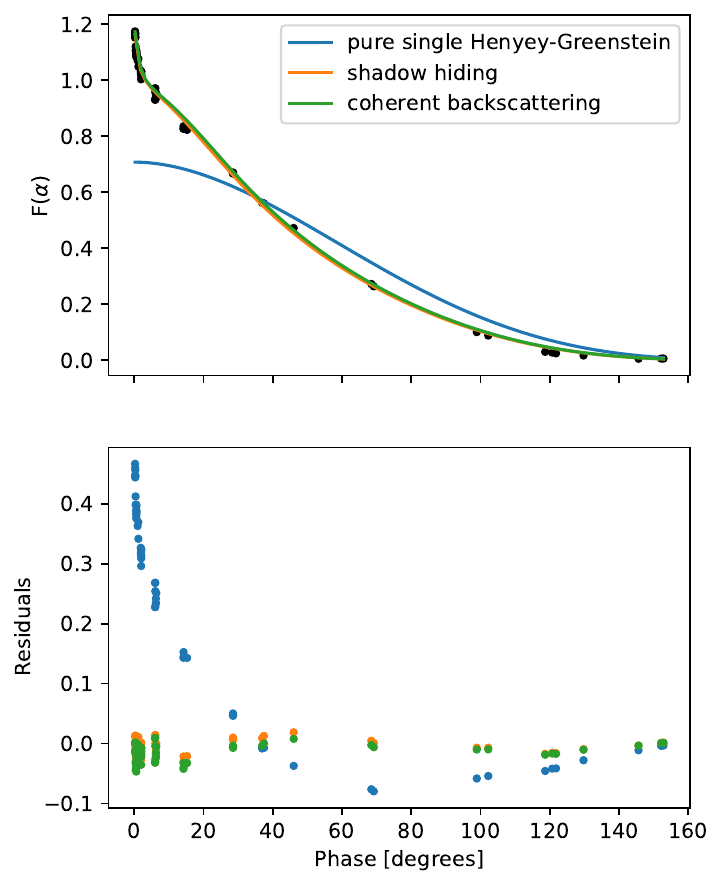}
      \caption{Best fit models to one of one of Enceladus' phase curves (filter 798\,nm) for three different models: a pure single Henyey-Greenstein model with no additional opposition peak, shadow hiding model, and coherent backscattering model. Top panel shows the fits and the second panel shows the residuals of these best fits. It is clear that the model with no opposition peak is the worst fit to the data. The other two models produce, by eye, almost identical fits, which is expected since the functional forms are similar. However the residuals still show subtle differences. Using the LOO model selection statistic allows us to conclude that the coherent backscattering model is preferred by 7/12 of the phase curves, however not significantly. The corner plot showing the posteriors for the shadow hiding model are in Appendix \ref{sec:bestfitresultsappendix}, Figure \ref{fig:enc_cornerplot}.}
      \label{fig:enc_bestfitmodels}
\end{figure}

Figure \ref{fig:enc_bestfitmodels} shows one of the Enceladus phase curves (observed with the 798\,nm filter), along with the best fits for the CB, SH, and no opposition peak models. We find that the models with an opposition peak were a much better fit for all phase curves. 

\begin{figure}
   \centering
   \includegraphics[width=9cm]{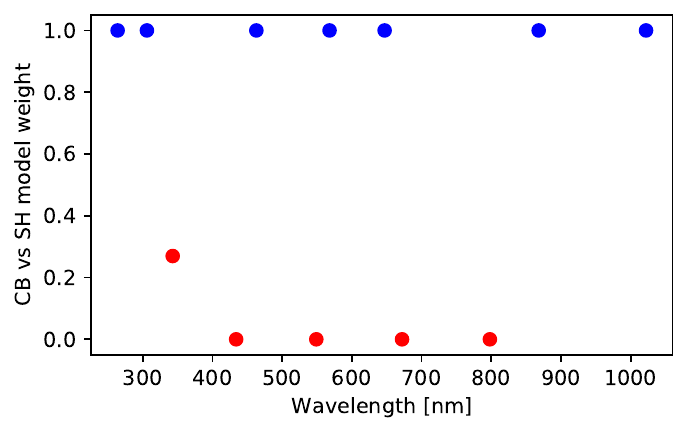}
      \caption{LOO model weight of the CB model vs the SH model for the Enceladus phase curves. When the model weight is close to 1, then the CB model is preferred (blue points), however close to 0 indicates that SH is preferred (red points). The red points show the phase curves where the SH model is confirmed. Bottom panel shows the LOO model weight plotted against the number of datapoints in each phase curve. There is a tentative negative correlation here, indicating that the more datapoints in a phase curve, the stronger the SH model is preferred. 
      }
         \label{fig:enc_wavelengthvselpddiff}
\end{figure}

Seven out of 12 Enceladus phase curves prefer CB, and the other five prefer SH (see Figure \ref{fig:enc_wavelengthvselpddiff}). Examining the top panel of Figure \ref{fig:enc_wavelengthvselpddiff}, we see no correlation with model selection and wavelength. The non-preference of either model here could suggest that the data does have sufficient precision to distinguish between them. It also does not rule out the possibility that both effects are present on Enceladus and the opposition peak that we observe is a complex combination of the two. For example, \cite{Dlugach2013} showed that CB can take place in closely packed mediums and planetary surfaces and, as mentioned previously, we also have evidence for both SH and CB on the Moon.

\begin{table}[h!]
\centering
\caption{Enceladus phase curves' preferred models}
\begin{tabular}{ |c|c|c|c| } 
 \hline
 Wavelength [nm] & Filter & Preferred Model & Weight \\
 \hline
 260 & UV1 & CB & 1.00\\ 
 310 & UV2 & CB& 1.00\\
 340 & UV3 & SH& 1.00\\
 430 & - & SH & 0.73\\
 460 &BL1 &  CB& 1.00\\
 550 & -  & SH& 1.00\\
 570 & GRN &   CB& 1.00\\ 
 650 & RED &  CB& 1.00\\
 670 & - & SH& 1.00\\
 800 & - & SH& 1.00\\
 870 & - & CB& 1.00\\
 1020 & - & CB& 1.00\\
 \hline
\end{tabular}
\tablefoot{For each of Enceladus' Cassini phase curves we report here the preferred model from our fitting procedure. 7/12 phase curves prefer the coherent backscattering model (CB) and the others prefer the shadow hiding model (SH). The wavelength column refers to the effective wavelength of the filter used. `Weight' is the LOO model weight of the preferred model.}
\label{tab:enc_modelselect}
\end{table}

\subsection{Information from the fitted parameters}

\begin{figure}
   \centering
   \includegraphics[width=\hsize]{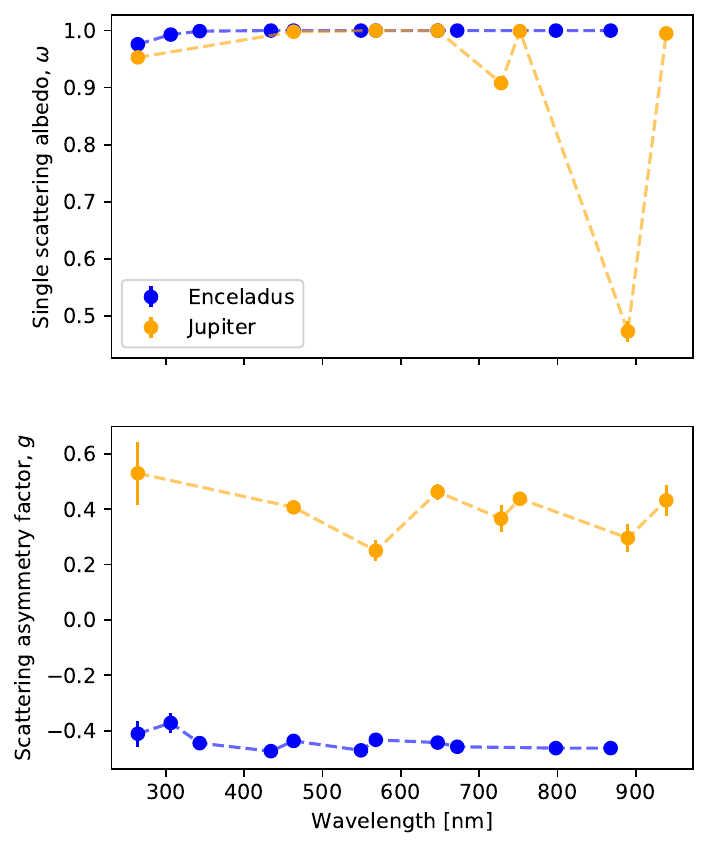}
      \caption{Fitted values of the scattering asymmetry factor, g, and the single-scattering cross section, $\omega$, across wavelength for Jupiter (orange) and Enceladus (blue). In the first panel, we see the single scattering albedo decrease (i.e. more absorption) around 750\,nm and 900\,nm for Jupiter, which is attributed to the methane absorption bands. We do not see this feature in Enceladus, since it has no considerable atmosphere. Both Jupiter and Enceladus show no trends in the asymmetry factors (second panel) across the wavelength range. This could imply, without taking into account surface roughness, that the size of the scattering particles are always equivalent to the wavelengths in this range, and therefore inhomogenous in these cases. For Enceladus, g is significantly lower, and negative, indicating that most of the light that hits the scattering particles (on the surface) is reflected backwards. This follows our intuition since Enceladus has an icy surface and no atmosphere, compared to Jupiter's thick atmosphere and no solid surface.}
         \label{fig:all_compare}
\end{figure}

\begin{table*}
\caption{Best-fit parameter results of the SH model fitting to Jupiter phase curves, including the priors used for the fitted parameters and the results of the derived parameters. }
\centering
\footnotesize
\scalebox{0.83}{
\begin{tabular}{lccccccccc}
{\bf Parameter} & {\bf Prior} & {\bf 260nm(UV1)} & {\bf 460nm(BL1)} & {\bf 570nm(GRN)} &{\bf 650nm(RED)}&{\bf 730nm(MT2)}&{\bf 750nm(CB2)}&{\bf 890nm(MT3)}&{\bf 940nm(CB3)}\\ \hline \\
\multicolumn{10}{l}{\bf \tiny Fitted Parameters} \\\\
g  &U(-1,1) & $0.5\pm0.1$ & $0.41\pm0.02$ & $0.25\pm0.04$ & $0.46\pm0.03$ & $0.37\pm0.05$ & $0.44\pm0.02$ & $0.30\pm0.05$ & $0.43\pm0.06$\\\\
$\omega$ & U(0,1) & $0.954\pm0.004$ & $0.9977\pm0.0002$ & $0.99998\pm0.00003$ & $0.9997\pm0.0001$ & $0.908\pm0.004$ & $0.9995\pm0.0001$ & $0.47\pm0.02$ & $0.995\pm0.001$\\\\
$h_{\textrm{SH}}$ [radians] & U(0,10) & $0.3\pm0.2$ & $0.18\pm0.01$ & $0.22\pm0.03$ & $0.10\pm0.02$ & $0.19\pm0.02$ & $0.06\pm0.01$ & $0.66\pm0.09$ & $0.21\pm0.05$\\\\
B0 & U(0,10) & $0.4\pm0.2$ & $0.56\pm0.02$ & $0.46\pm0.04$ & $0.26\pm0.02$ & $0.51\pm0.03$ & $0.155\pm0.009$ & $3.1\pm1.2$ & $0.37\pm0.06$\\\\ \hline \\
\multicolumn{10}{l}{\bf \tiny Derived Parameters} \\\\
FWHM [radians] & & $0.7\pm0.4$ & $0.35\pm0.02$ & $0.44\pm0.07$ & $0.20\pm0.04$ & $0.39\pm0.03$ & $0.17\pm0.02$ & $1.3\pm0.2$ & $0.4\pm0.1$  \\\\
\end{tabular}}
\label{tab:bigtable_jup}
\end{table*}

\begin{table*}
\caption{Best-fit parameter results of the SH model fitting to Enceladus phase curves, including the priors used for the fitted parameters and the results of the derived parameters. }
\centering
\footnotesize
\scalebox{0.65}{
\begin{tabular}{lcccccccccccc}
{\bf Parameters} & {\bf Prior} & {\bf 260nm(UV1)} & {\bf 310nm(UV2)} & {\bf 340nm(UV3)} & {\bf 430nm} & {\bf 460nm(BL1)} & {\bf 550nm} & {\bf 570nm(GRN)} &{\bf 650nm(RED}&{\bf 670nm}&{\bf 800nm}&{\bf 870nm}\\ \hline \\
\multicolumn{13}{l}{\bf \tiny Fitted} \\\\
g  &U(-1,1) & $-0.41\pm0.05$ & $0.37\pm0.04$ & $-0.445\pm0.008$& $-0.474\pm0.009$& $-0.437\pm0.006$& 
$-0.471\pm0.007$& $-0.433\pm0.007$& $-0.443\pm0.01$& $-0.458\pm0.007$& $-0.463\pm0.005$ & $-0.463\pm0.008$\\\\
$\omega$ & U(0,1) & $0.98\pm0.01$ & $0.993\pm0.05$ & $0.999\pm0.001$& $1.000\pm0.000$& $1.000\pm0.000$& $1.000\pm0.000$& $1.000\pm0.000$& $1.000\pm0.000$& $1.000\pm0.000$& $1.000\pm0.000$& $1.000\pm0.000$\\\\
$h_{\textrm{SH}}$ [radians] & U(0,10) & $0.2\pm0.8$ & $0.07\pm0.05$ & $0.007\pm0.002$ & $0.0090\pm0.004$ & $0.017\pm0.006$ & $0.007\pm0.002$ & $0.016\pm0.008$ & $0.03\pm0.02$ &$0.013\pm0.003$ & $0.009\pm0.002$ & $0.008\pm0.003$\\\\
B0 & U(0,10) & $0.7^{+1.4}_{-0.6}$ & $0.5\pm0.2$ & $0.46\pm0.03$ & $0.26\pm0.02$ & $0.24\pm0.01$ & $0.28\pm0.01$ & $0.21\pm0.01$ & $0.27\pm0.02$ & $0.027\pm0.01$ & $0.298\pm0.009$ & $0.29\pm0.02$\\\\ 
\hline \\
\multicolumn{13}{l}{\bf \tiny Derived} \\\\
FWHM [radians] & & $0.4^{+1.7}_{-0.3}$ & $0.1\pm0.1$ & $0.014\pm0.004$ & $0.018\pm0.008$ & $0.03\pm0.01$ & $0.014\pm0.004$ & $0.03\pm0.02$ & $0.06\pm0.04$ & $0.026\pm0.006$ & $0.018\pm0.004$ & $0.016\pm0.006$  \\\\
\end{tabular}}
\label{tab:bigtable_enc}
\end{table*}

The best-fit results of the fitted parameters for Jupiter and Enceladus (with the shadow-hiding model) are shown in Tables \ref{tab:bigtable_jup} and \ref{tab:bigtable_enc}. The single-scattering albedo $\omega$ describes the fraction of light reflected in a single scattering event. When $\omega$ is close to unity, the majority of the light is scattered, and absorption dominates as $\omega \rightarrow 0$. In the top panel of Figure \ref{fig:all_compare}, we find $\omega\sim1$ for Enceladus across the full range of wavelengths. This is consistent with Enceladus having no atmosphere to absorb the incoming light and an icy solid surface which has a high optical albedo. Jupiter also has $\omega\sim1$ for most of the wavelength range, apart from methane absorption bands near 750\,nm and 900\,nm.

The scattering asymmetry parameter, $g$, parametrises a bias towards forward- or backscattering (see Equation \ref{eqn:shg}). Forward scattering dominates for positive $g$, and backward scattering dominates for negative $g$. The fit results for Jupiter show $g > 0$, indicating that the scattering surface is composed of particles larger than the observing wavelengths \citep{Heng2021a}.  That $g$ remains relatively constant across the wavelength range could imply the presence of a size distribution of particles \citep{Heng2021a}. For Enceladus, the inferred values for $g$ are negative implying that reverse scattering dominates. This follows our intuition since Enceladus has an icy surface and no atmosphere, compared to Jupiter's thick atmosphere and no solid surface. However, due to the single Henyey-Greenstein scattering phase function being developed to describe scattering off a particle, and not reflection off a surface, it is less clear how to interpret $g$ in the context of a solid surface.

From the inferred values of $h_{\rm CB}$ (not shown), the transport mean free path associated with coherent backscattering is $\sim 0.1$--1 times the wavelength probed.

\subsection{FWHM as a solid surface detector}
\label{sec:fwhm}
\begin{figure}
   \centering
   \includegraphics[width=\hsize]{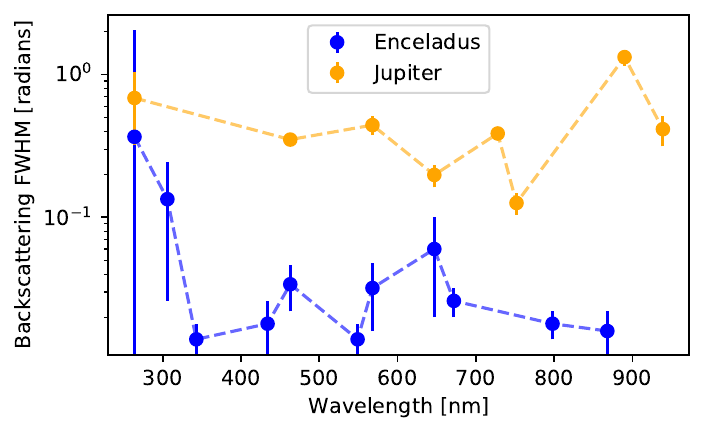}
      \caption{Measured backscattering FWHM in the Jupiter (orange points) and Enceladus (blue points) phase curves. We find that there is generally an order of magnitude difference between the Enceladus and Jupiter measurements across the wavelength range.}
         \label{fig:fwhm}
\end{figure}

From previous studies \citep[e.g.][]{Dyudina2016}, we see that one of the distinguishing features between the opposition effect from our Solar System objects is the `peaky-ness' of the opposition maximum near zero phase. To quantify this, we used the best fit SH model to measure the FWHM (Equation \ref{eqn:fwhm}) of the different Jupiter and Enceladus phase curves and plotted them as a function of wavelength (see Figure \ref{fig:fwhm}).

We find the FWHM of Enceladus' opposition peak to be an order of magnitude larger than Jupiter across the wavelength range. This helps quantify what we have seen in the rest of the Solar System, which is that the solid planets seem to have much smaller opposition FWHM peaks than the gaseous planet \citep{Dyudina2016}. This could be due to the difference in how light is scattered on a surface compared to within a gas. It could also be due to the dominant backscattering source or even the average distance between the scattering particles themselves.

\section{Application to exoplanet phase curves}
\subsection{Rocky exoplanets}
\label{sec:rockyexoplanets}
Characterising the climates of exoplanets is now one of the most important topics of research in the exoplanet field. Finding more methods to do this is therefore extremely valuable. We have so far proposed the possibility that measuring an opposition peak in the phase curve of a planet and then measuring the FWHM of this peak would be an indicator of whether the planet has a solid or gaseous surface. 

\begin{figure}
   \centering
   \includegraphics[width=\hsize]{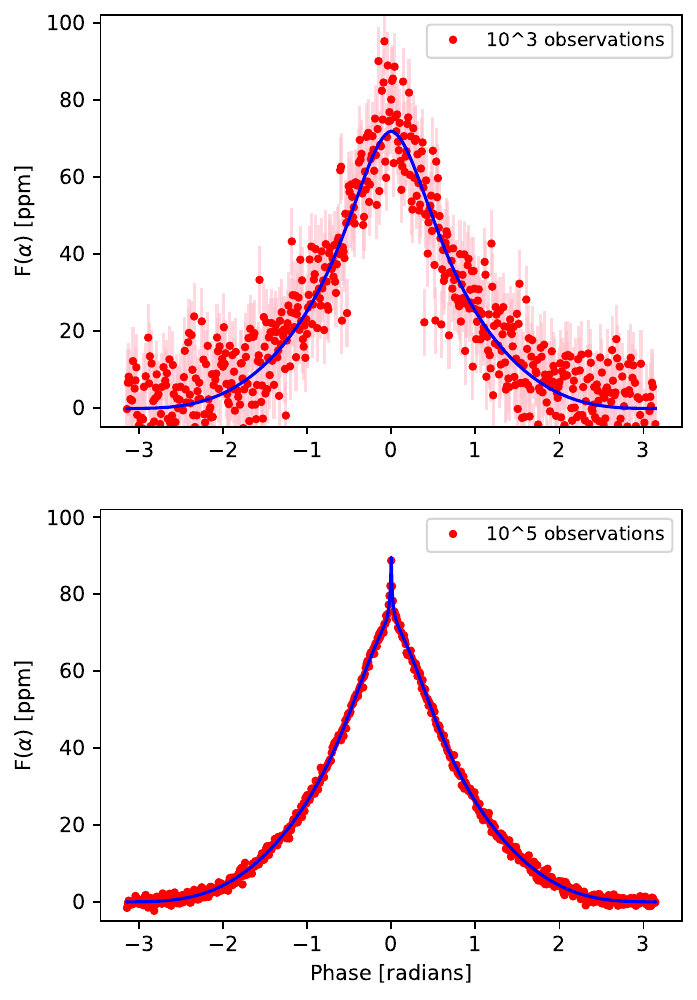}
      \caption{Simulated phase curve model of K2-141b, a super-Earth (red points, with green errorbars). The blue line shows the best fit of our opposition effect model. The noise here is at a level consistent with $10^3$ JWST observations (top panel) and $10^5$ observations (bottom panel).}
         \label{fig:fittingexample}
\end{figure}

\begin{figure}
   \centering
   \includegraphics[width=\hsize]{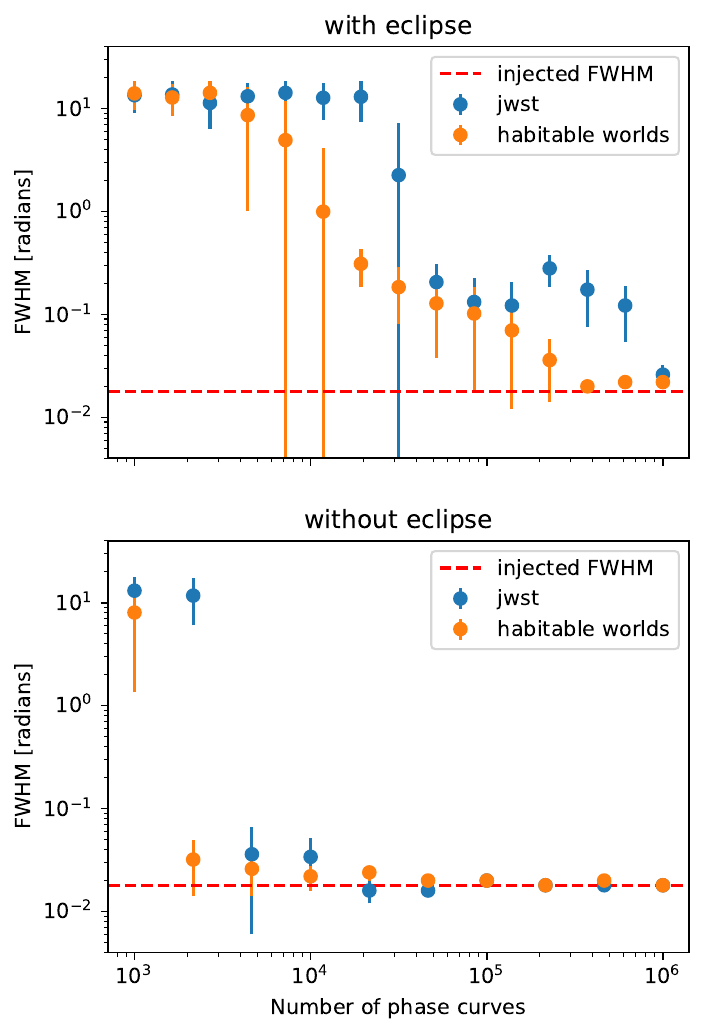}
      \caption{Feasibility of detecting an Enceladus-like opposition effect signal in an exoplanet phase curve using both JWST and the future Habitable Worlds Observatory. The first panel shows the results when we use phase curves with an eclipse present and the second panel shows results without an eclipse present. The red line shows the FWHM value of the injected signal and the blue and orange points show the retrieved value, with $1\sigma$ errorbars. We find that, as expected, this signal is easier to detect in phase curves without an eclipse present, however this is still way beyond feasibility for both JWST and Habitable Worlds.}
         \label{fig:fwhm_jwst}
\end{figure}

In order to test whether this would be possible with an exoplanet, we ran a series of injection-recovery tests. Using a simulated phase curve of K2-141b, one of the most observable super-Earths \citep[see, e.g.][]{Barragan2018, Malavolta2018}, we injected an Enceladus-like opposition peak using the best-fit parameters from our model fitting in Section \ref{sec:fwhm}. We took the SH model results since they were not ruled out by the data and as we only want to measure the FWHM of the peak, both the CB and SH fit should give the same result. Additionally, the formula for calculating the FWHM from the SH is easily given by Equation \ref{eqn:fwhm}. We then adjusted the noise floor to simulate observations from both JWST and a future Habitable Worlds Observatory-like telescope. We took performance specifications from \citet{TheLUVOIRTeam2019} as a starting point for the Habitable Worlds Observatory, and in-flight spectrophotometric precision measurements for JWST/NIRSpec from \citet{Mikal-Evans2023}. We investigated how many observations are necessary to measure the injected (true) FWHM. Figure \ref{fig:fittingexample} shows the simulated phase curve (without the eclipse) with noise levels consistent with $10^3$ and $10^5$ JWST observations. From these fits we obtain the $h_{\textrm{SH}}$ posterior distribution, allowing us to calculate the FWHM as shown in Figure \ref{fig:fwhm_jwst}. 

Our results are shown in Figure \ref{fig:fwhm_jwst}. We ran the test with two different simulated phase curves: one with and one without an eclipse present. This was important for us to test since an eclipse blocks the light from the planet near zero phase, exactly when the opposition peak signal is the strongest. Therefore we expect a present eclipse to make the fits less accurate than when we have all the data near zero phase. A phase curve with no eclipse could be produced by a non-transiting planet.

From our results we see that with JWST and a phase curve with an eclipse, it  would take upwards of $\sim10^6$ phase curves in order to reach the precision necessary to reach FWHM measurements consistent with the true value. Using the lower predicted noise floor of Habitable Worlds, this decreases to $\sim10^5$ phase curves. Comparing these to the results when we use a phase curve with no eclipse present, the number of required observations significantly decreases. JWST requires $\sim10^4$ phase curves, whilst Habitable Worlds requires $\sim10^3$. Clearly both of these results show that this technique is not feasible with both current and planned future individual exoplanet observations. Perhaps it could still be used in the future for non-transiting exoplanets if a global population phase curve stacking technique was applied \citep[e.g.][]{Sheets2017}. 

\subsection{Jupiter-like exoplanets}
We repeated the above injection and recovery test using simulated phase curves of HAT-P-7b and injecting them with an opposition-effect signal using the best fit parameters from the previous analysis on the Jupiter phase curves. We then used an MCMC fitting method to recover the $h_{\textrm{SH}}$ parameter (and therefore the FWHM) and compared this to the true value of the injected FWHM.

We find that both with and without an eclipse present in the phase curve, the number of observations required to contrain the FWHM is $\sim10^2$ lower than our previous investigation with a rocky exoplanet.

Since the FWHM we injected is much wider than the one we injected to mimic a rocky exoplanet, it has an effect over a larger range of phases, so more data contributes to constraining the opposition effect parameters, and fewer observations are needed. The second is that the orbital period is longer, so at a fixed exposure cadence, there are more observations per phase. Furthermore, the $R_p/a$ for this type of planet is much larger than for the Enceladus-like planet (mostly due to the larger radius), increasing the amplitude of the phase curve and therefore our S/N. In Section \ref{sec:optimalparams}, we investigated the parameter space of planetary orbital period and semi-major axis to see where the opposition signal would have the highest signal-to-noise (S/N).  

\section{Discussion}
\label{sec:discussion}
\subsection{The effect of neglected surface properties on reflectance}
The reflected light phase curve model in Section~\ref{sec:theory} describes a spherical planet built on formalism established in \citet{Hapke1981} and extended in \citet{Heng2021b}. We adopted a parametrisation for shadow hiding from \citet{Hapke2002}, which implied that shadows are cast by topographical features that could be any optically thick obstruction to light, including mountains or clouds. The scale and shape of the shadow-casters was not parametrised in the model.

The full diversity of planetary surfaces in the Solar System cannot be described with shadow hiding or coherent backscattering, alone or in combination. \citet{Hapke1984} established a parametrisation for reflected light phase curves with macroscopic roughness. Model lunar phase curves with surface roughness agreed well with observations after taking into account the distribution of surface slopes measured from lunar orbiters \citep{Helfenstein1988}. Later work also established the effect of regolith porosity on phase curves \citep{Hapke2008}. 

A primary motivation for this study was to evaluate whether or not opposition surges are useful solid surface detectors for exoplanets. In Section~\ref{sec:rockyexoplanets}, we showed that Enceladus-like opposition surges are unlikely to be measured in exoplanet phase curve observations in the foreseeable future. As such, we have not added further complexity to the models fit to the phase curves of Enceladus or Jupiter, since it is unlikely to affect the likelihood of opposition surge detections in exoplanets. Much higher precision phase curves are available elsewhere, and their precision requires the inclusion of these higher order effects in model fits. For example, Enceladus observations from Voyager and from the ground by \citet{Verbiscer1994} required macroscopic roughness to model the phase curve. Since the phase curve model here is parametrised through the single scattering albedo and scattering asymmetry parameter, the inferences for both parameters on Enceladus must be interpreted with these model limitations in mind. However, in, for example, \cite{Buratti1985}, they found that topographical effects are not expected to be significant at phase angles less than $30-40^{\circ}$, angles significantly larger than our focus for this study.

Despite our highly simplified model and wider phase angle coverage, the results presented in this work agree well with \citet{Verbiscer1994}. \citet{Verbiscer1994} found $\omega=0.999 \pm 0.002$ and $g = -0.39 \pm 0.02$, with opposition surge factor $B_0 = 0.21 \pm 0.07$. The Voyager instrument response with the clear filter has most transmittance between 400-600 nm, so the Cassini filter at 550 nm is likely the closest analogue in Table~\ref {tab:bigtable_enc}. We find smaller $g$ ($-0.471\pm0.007$) and greater $B_0$ ($0.28 \pm 0.01$) in the Cassini 550 nm filter than the Voyager clear filter observations. The difference between solutions may be attributable to the difference in filter transmittance or the inclusion of regolith compaction in the analysis by \citet{Verbiscer1994}. The small disagreement between parameters inferred in the two analyses is insufficient to change the outcome of the likelihood of opposition surge detection in exoplanets in Section~\ref{sec:rockyexoplanets}.

\subsection{Selecting an appropriate model}

We fitted a set of reflected light phase curve models to Solar System observations. The scattering phase functions of these models are semi-empirical and constructed to fit observations of Solar System objects, but it is not clear from the theory which models should be applied to a given planet. Given the similarity of the model shapes, we have seen that trying to distinguish the preferred model (CB or SH) from the observations used in this study is very difficult. We could only tentatively conclude there was evidence for CB model being preferred for Jupiter. This of course does not rule out the possibility of both effects being present with one more dominant than the other or that the true solution is a different model entirely. For example, in \cite{Verbiscer2005}, they found that for HST observations of Enceladus between 338-1022nm, their best model was a combination of CB and SH, something that our work in Section \ref{sec:modelcomp} does not disagree with. However, as previously mentioned, our model does differ slightly from the model employed in \cite{Verbiscer2005} (see Section \ref{sec:theory} for an overview of our model) so comparing these directly should be done with caution. Even for the Moon, where studies have been made on the Moon rock samples themselves, there are still many outstanding questions in regards to the contributions of CB and SH to the lunar opposition peak. There have been studies \citep[see, e.g.][]{Helfenstein1997} showing evidence that CB contributes to the peak at very small phase angles, and SH continues out to larger phase angles - however this is not conclusive. There is also very little investigation into the cause of the observed opposition effect on gaseous planets. 

Furthermore, due to the known wavelength-dependence of some scattering properties, we expected some colour-dependence in the best-fit parameters however this was not detected in our results. This could be due to the limited (and mostly optical) wavelength range where the albedo of these objects is both high. It has been seen in \cite{Buratti2022} that due to SH and CB acting differently on singly and multiply scattered light, the opposition surge form changes as you go to IR wavelengths. They hypothesise that this is due to the lower IR albedo reducing the influence of the multiply scattered light. However, due to our model formulation and since the forms for SH and CB are empirical and very similar (see, e.g. Section \ref{sec:modelcomp}), the data we use is not precise enough to separate them and comment on this prediction.

However, due to SH and CB being direct outcomes of the scattering properties of the surface and/or atmosphere, unravelling these dependencies would provide information about the composition and particles on these planets.

\subsection{Interpreting the best-fit parameters}

For fitting our reflected light curves, we used, like other previous studies \citep[see, e.g.][]{Helfenstein1997, Heng2021a}, a single Henyey-Greenstein scattering phase function to describe the preferred scattering direction of an incoming ray to a scattering particle. This function includes the scattering asymmetry parameter, $g$, which controls the amount of forward vs backward backscattering. When $g$ is close to 1 there is only forward scattering, whilst if $g$ is close to -1 then there is only backscattering present. Fitting for Enceladus, we derived a negative $g$. This is more difficult to interpret since the scattering medium of Enceladus is a surface and not atmospheric gas. Therefore $g$ is harder to interpret in this scenario, which is not the medium the phase function was designed to model. Also, the scattering properties that can be derived from $g$ are probably not reliable for describing Enceladus' suface. For Jupiter, since it has a large extended atmosphere, we believe this model is appropriate. Future works should take care in choosing the scattering phase function and be aware of its limitations in regards to solid bodies without atmospheres. 

\subsection{Optimal planet parameters for detecting the opposition effect}
\label{sec:optimalparams}

For a fixed planet radius $R_p$ and opposition scattering parameters, two parameters dominate the S/N of the opposition peak: planetary orbital period and stellar mass. As the period increases, the number of datapoints per phase increases by the same amount. Therefore the S/N due to the number of datapoints goes as
\begin{equation}
\rm{S/N} \propto N_{\rm{exp}}^{1/2} \propto P^{1/2},
\end{equation}
where $N_{\rm{exp}}$ is the number of exposures and P is the planetary orbital period. As the mass of the host star increases, by Kepler's Third Law, the semi-major axis ($a$) also increases, for a fixed period. The amplitude of the reflected light phase curve of a planet is equal to $A_g (R_p/a)^2$. Therefore the S/N goes as
\begin{equation}
\rm{S/N} \propto \rm{amplitude} \propto a^{-2} \propto P^{-4/3}.
\end{equation}

These show that the amplitude of the phase curve dominates how the S/N of the opposition peak changes with period. That is, as the period of the planet decreases, the S/N increases.

\section{Conclusion}
\begin{itemize}
    \item Reflected light phase curves of Solar System objects contain opposition peaks, which contain information about the surfaces where scattering occurs.
    \item We fit Jupiter and Enceladus phase curves with a semi-empirical reflected light phase curve model and found that models with an opposition peak model are significantly preferred over simpler models, in agreement with previous Solar System work.
    \item We showed that the FWHM of Jupiter's opposition peak is an order of magnitude larger than that of Enceladus, uncovering the opportunity to differentiate solid from gaseous exoplanets using the FWHM of phase curve opposition peaks.
    \item Cross-validation suggests a tentative preference for CB in Jupiter's phase curve over SH, and a preference for SH in the phase curve of Enceladus.
    \item We show that observations are unlikely to accurately measure the FWHM of a phase curve opposition peak in the next few decades, with either JWST or HWO. 
\end{itemize}

\bibliographystyle{aa} 
\bibliography{bibliography} 

\appendix

\section{Best fit results}
\label{sec:bestfitresultsappendix}
\begin{figure}
   \centering
   \includegraphics[width=\hsize]{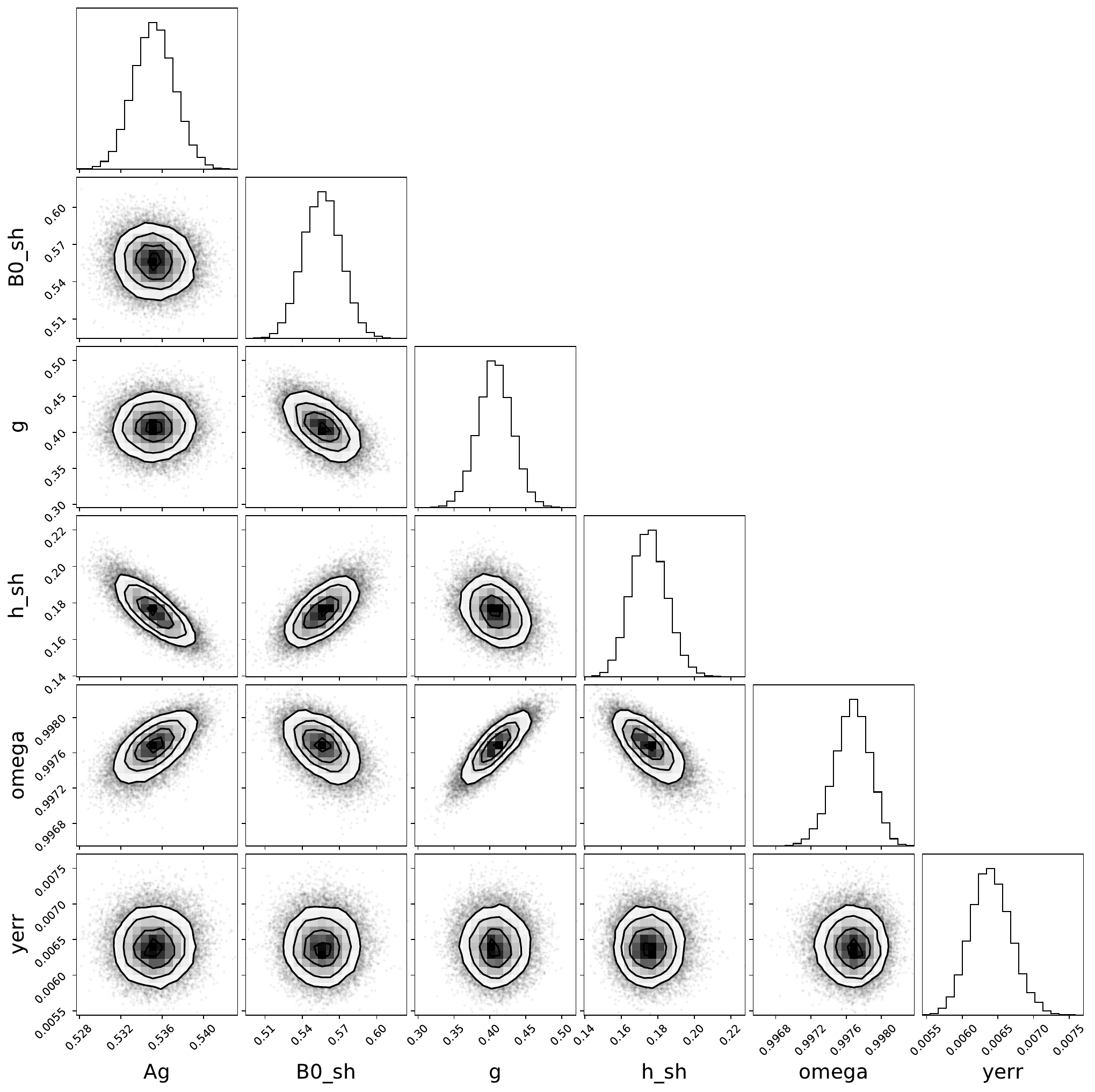}
      \caption{Corner plot showing the shadow hiding model best fit posteriors to the BL1 (463\,nm) Jupiter phase curve, as shown in Figure \ref{fig:bestfitmodels}.}
         \label{fig:jup_cornerplot}
\end{figure}

\begin{figure}
   \centering
   \includegraphics[width=\hsize]{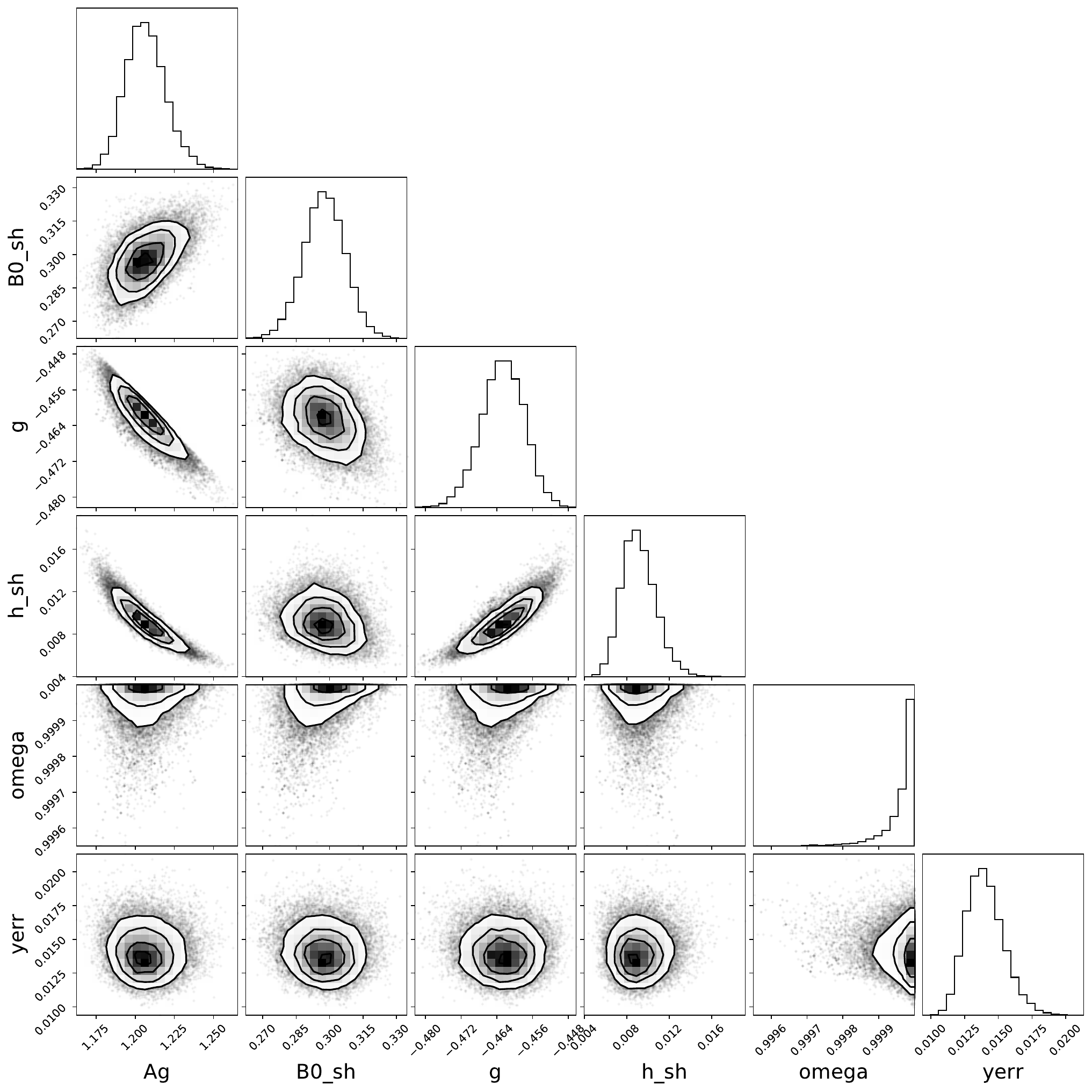}
      \caption{Corner plot showing the shadow hiding model best fit posteriors to the 798nm Enceladus phase curve, as shown in Figure \ref{fig:enc_bestfitmodels}.}
         \label{fig:enc_cornerplot}
\end{figure}

We show here in Figure \ref{fig:jup_cornerplot} and \ref{fig:enc_cornerplot} the resulting posterior distributions of our fitted parameters during our analysis in Section \ref{sec:modelcomp}. 

\section{Comparison to \cite{Heng2021a}}
This current analysis follows on from the work in \cite{Heng2021a}, where they tested different single scattering laws on a similar dataset from \cite{Li2018} of Jupiter phase curves. These phase curves not only consisted of observed data, but were interpolated so that they had homogeneous phase and wavelength coverage. We initially repeated our analysis in Section \ref{sec:modelcomp} with these interpolated phase curves. However we found this often led to bimodal solutions, which we speculated was an artifact of the interpolation procedure. Therefore we decided to use the original uninterpolated data in this analysis.  

\end{document}